\documentclass{PoS}

\newcommand{\dhd}{{\textstyle d}
\lower.03ex\hbox{\kern-0.40em$^{\scriptstyle-}$}\kern-0.08em{}}

\newcommand{\calf}{{\cal F}}
\newcommand{\tilcaf}{\tilde{\cal F}}
\newcommand{\ticalf}{\tilde{\cal F}}
\newcommand{\cald}{{\cal D}}
\newcommand{\bu}{{\bullet}}
\newcommand{\tilF}{\tilde{F}}
\newcommand{\calh}{{\cal H}}
\newcommand{\tilU}{\tilde{U}}
\newcommand{\half}{{1\over 2}}
\newcommand{\calo}{{\cal O}}
\newcommand{\calr}{{\cal R}}

\title{Evolution of gluon TMDs from small to moderate x}

\ShortTitle{Evolution of gluon TMDs from small to moderate x}

\author{Andrey Tarasov\\
        Theory Group, Jefferson Lab (JLAB), 12000 Jefferson Ave, Newport News, VA 23606,USA\\
        E-mail: \email{atarasov@jlab.org}}

\abstract{Recently we obtained an evolution equation of gluon TMDs, which addresses a problem of unification of different kinematic regimes. It describes evolution in the whole range of Bjorken $x_B$ and the whole range of transverse momentum $k_\perp$. In this notes I study different limits of this evolution equation and show how it yields several well-known and some previously unknown results.}

\FullConference{QCD Evolution 2015 -QCDEV2015-\\
		26-30 May 2015\\
		Jefferson Lab (JLAB), Newport News, Virginia, USA}

\begin{document}

\section{Introduction}
A concept of factorization plays an important role in study of high-energy scattering reactions. It implies that parts of interaction that correspond to different scales can be separated from each other. The scheme of this separation is not unique and depends on the process and its kinematic regime. 
In particular, for analysis of semi-inclusive reactions like semi-inclusive deep inelastic scattering (SIDIS), Drell-Yan, $e^+e^-$ annihilation, etc. one can apply transverse momentum dependent (TMD) factorization \cite{cs1, jimayuan, collinsbook}. In this case parton distribution functions (PDFs) depend both on longitudinal momentum fraction variable $x$ and transverse momentum $k_\perp$ of the parton involved.

A great success has been achieved in understanding of TMD factorization at moderate-$x$ \cite{cs1, jimayuan, collinsbook, echevidsci,muldrod, cs2}. At the same time, for small-$x$ a variety of different approaches has been developed \cite{domarxian, chixiyu, muxiyu, kovsievert}. This approaches are rather different from their moderate-x counterparts. As a result one should address a question how to describe transition between two limits and get some kind of unified picture.

Of course, one can argue that "variations" between different approaches are not relevant, because each of them is justified in its own kinematic region and in a sense doesn't interfere with another. However, while in theory the regimes are well separated, in practice their boarders are not rigorously defined. For example, the future Electron Ion Collider (EIC) \cite{EIC} will work at intermediate values of $x$ and, while being a moderate-$x$ machine, will reveal a spectrum of phenomena related to small-$x$ physics. This indicates the need to study a transition from low to moderate-$x$. This of course would be very interesting from theoretical point of view as well.

It was shown recently that such transition can be constructed \cite{mpaper, mpaper2}. We consider a rapidity evolution of TMD operator which generates a set of (un)polarized TMD PDFs:
\begin{eqnarray}
\ticalf_i^a(\beta_B,x_\perp)\calf_j^a(\beta_B,y_\perp),
\label{operator}
\end{eqnarray}
where
\begin{eqnarray}
&&\hspace{-0mm} 
\calf^{a\eta}_{i}(\beta_B,k_\perp)~=~\!\int\! d^2z_\perp~e^{-i(k,z)_\perp}\calf^{a\eta}_{i}(\beta_B,z_\perp),
\nonumber\\
&&\hspace{-0mm} 
\calf^{a\eta}_{i}(\beta_B,z_\perp)~\equiv~{2\over s}
\!\int\! dz_\ast ~e^{i\beta_B z_\ast} 
\big([\infty,z_\ast]_z^{am}gF^m_{\bu i}(z_\ast,z_\perp))^\eta
\label{kalf}
\end{eqnarray}
and
\begin{eqnarray}
&&\hspace{-0mm} 
\tilcaf_i^{a\eta}(\beta_B,k_\perp)~=~\!\int\! d^2z_\perp~e^{i(k,z)_\perp}\tilcaf_i^{a\eta}(\beta_B,z_\perp),
\nonumber\\
&&\hspace{-0mm} 
\tilcaf^{a\eta}_i(\beta_B,z_\perp)~\equiv~{2\over s}
\!\int\! dz_\ast ~e^{-i\beta_B z_\ast} 
g\big(\tilF^m_{\bu i}(z_\ast,z_\perp)[z_\ast,\infty]_z^{ma}\big)^\eta.
\label{tilkalf}
\end{eqnarray}
We use the Sudakov decomposition $k=\alpha p_1+\beta p_2+k_\perp$, where $p_1$ and $p_2$ are two light-like vectors defined by the process. For the coordinates we use 
the notations $x_\bu\equiv x_\mu p_1^\mu$ and $x_\ast\equiv x_\mu p_2^\mu$ related to the light-cone coordinates by $x_\ast=\sqrt{s\over 2}x_+$ and $x_\bu=\sqrt{s\over 2}x_-$. We study a case of future-point Wilson lines defined as
\begin{eqnarray}
&&\hspace{-0mm} 
 [x_\ast, y_\ast]^\eta_{z_\perp}~=~{\rm Pexp}\Big[\frac{2ig}{s}\!\int_{y_\ast}^{x_\ast}\!\! dz_\ast ~p_1^\mu A^\eta_\mu(z_\ast, z_\perp)\Big].
\end{eqnarray}

The tilde sign indicates that the corresponding operators obey an inverse time ordering. Variable $\beta_B$ serves as a longitudinal momentum fraction $x$ for distribution functions and $1/z_F$ for fragmentation.

In this study we use a rapidity factorization approach \cite{npb96}. We introduce a rapidity cutoff $\sigma$ and separate all fields into two classes based on their Sudakov $\alpha$. As a results we can write a particular semi-inclusive cross section as a convolution of the hard part (impart factor) constructed from  "slow" fields with $\alpha>\sigma$ and the matrix element of operator (\ref{operator}) constructed of "fast fields"
\begin{eqnarray}
&&\hspace{-0mm}
A^\eta_\mu(x)~=~\int\!{d^4 k\over 16\pi^4} ~\theta(e^\eta-|\alpha|)e^{-ik\cdot x} A_\mu(k),
\label{cutoff}
\end{eqnarray}
where $\eta\equiv\ln\sigma$.

To get the evolution equation of operator (\ref{operator}) we study its dependence on the rapidity cut-off. We shift the cut-off to a new value $\sigma'$ and calculate one-loop corrections to (\ref{operator}) with the gluon emission within $\sigma'<\alpha<\sigma$ in the background of the target fields (gluons with $\alpha<\sigma'$). The rapidity divergence is regularized by explicit cut-off $\sigma$. We calculate an evolution kernel in moderate- and small-x limits and then combine two results. 

One can find details of derivation in Ref. \cite{mpaper}. The final evolution equation for the matrix element of (\ref{operator}) between two proton states is
\begin{eqnarray}
&&\hspace{-1mm}
{d\over d\ln\sigma}\langle p|\tilcaf_i^a(\beta_B, x_\perp) \calf_j^a(\beta_B, y_\perp)|p\rangle~
\label{masterdis}\\
&&\hspace{-1mm}
=~-\alpha_s\langle p|{\rm Tr}\Big\{\!\int\!\dhd^2k_\perp\theta\big(1-\beta_B-{k_\perp^2\over\sigma s}\big)\Big[
(x_\perp|\Big( U{1\over\sigma\beta_Bs+p_\perp^2}
( U^\dagger k_k+p_k U^\dagger)
\nonumber\\
&&\hspace{-1mm}
\times~{\sigma \beta_Bsg_{\mu i}-2k^\perp_{\mu}k_i\over\sigma \beta_Bs+k_\perp^2}-~2k^\perp_\mu g_{ik} U{1\over \sigma\beta_Bs+p_\perp^2} U^\dagger
-2g_{\mu k}  U{p_i\over\sigma\beta_Bs+p_\perp^2} U^\dagger\Big)
\ticalf^k\big(\beta_B+{k_\perp^2\over\sigma s}\big)|k_\perp)
\nonumber\\
&&\hspace{-1mm}
\times~(k_\perp|\calf^l\big(\beta_B+{k_\perp^2\over\sigma s}\big)
\Big({\sigma \beta_Bs\delta^\mu_j-2k_\perp^{\mu}k_j\over\sigma\beta_Bs+k_\perp^2}(k_lU+Up_l){1\over\sigma \beta_Bs+p_\perp^2}U^\dagger
\nonumber\\
&&\hspace{-1mm}-2k_\perp^\mu g_{jl}U{1\over \sigma\beta_Bs+p_\perp^2}U^\dagger
-~2\delta_l^\mu U{p_j\over\sigma\beta_Bs+p_\perp^2}U^\dagger\Big)|y_\perp)
\nonumber\\
&&\hspace{-1mm}
+~2(x_\perp|\ticalf_i\big(\beta_B+{k_\perp^2\over\sigma s}\big)|k_\perp)
(k_\perp|\calf^l\big(\beta_B+{k_\perp^2\over\sigma s}\big)
\Big({k_j\over k_\perp^2}{\sigma \beta_Bs+2k_\perp^2\over\sigma\beta_Bs+k_\perp^2}(k_lU+Up_l)
{1\over\sigma \beta_Bs+p_\perp^2}U^\dagger
\nonumber\\
&&\hspace{-1mm}
+~2U{g_{jl}\over \sigma\beta_Bs+p_\perp^2}U^\dagger-2{k_l\over k_\perp^2}U{p_j\over \sigma\beta_Bs+p_\perp^2}U^\dagger
\Big)
|y_\perp)
\nonumber\\
&&\hspace{-1mm}
+~2(x_\perp|\Big( U{1\over\sigma\beta_Bs+p_\perp^2}
( U^\dagger k_k+p_k U^\dagger){k_i\over k_\perp^2}
{\sigma \beta_Bs+2k_\perp^2\over\sigma \beta_Bs+k_\perp^2}
+2 U{g_{ik}\over \sigma\beta_Bs+p_\perp^2} U^\dagger
\nonumber\\
&&\hspace{-1mm}
-~2 U{p_i\over\sigma\beta_Bs+p_\perp^2} U^\dagger{k_k\over k_\perp^2}\Big)
\ticalf^k\big(\beta_B+{k_\perp^2\over\sigma s}\big)|k_\perp)(k_\perp|\calf_j\big(\beta_B+{k_\perp^2\over\sigma s}\big)|y_\perp)\Big]
\nonumber\\
&&\hspace{-1mm}
+~2\ticalf_i(\beta_B, x_\perp)
(y_\perp|-{p^m\over p_\perp^2}\calf_k(\beta_B)(i\!\stackrel{\leftarrow}{\partial}_l+U_l)(2\delta_m^k\delta_j^l-g_{jm}g^{kl})
U{1\over \sigma\beta_Bs+p_\perp^2}U^\dagger|y_\perp)
\nonumber\\
&&\hspace{-1mm}
+~2(x_\perp|
 U{1\over \sigma \beta_Bs+p_\perp^2} U^\dagger(2\delta_i^k\delta_m^l-g_{im}g^{kl} )(i\partial_k- U_k)\tilde{\calf}_l(\beta_B)
{p^m\over p_\perp^2}|x_\perp)\calf_j(\beta_B, y_\perp)
\nonumber\\
&&\hspace{-1mm}
-~4\!\int\!{\dhd^2k_\perp\over k_\perp^2}\Big[\theta\big(1-\beta_B-{k_\perp^2\over\sigma s}\big)
\ticalf_i\big(\beta_B+{k_\perp^2\over\sigma s}, x_\perp\big)
\calf_j\big(\beta_B+{k_\perp^2\over\sigma s}, y_\perp\big)e^{i(k,x-y)_\perp}
\nonumber\\
&&\hspace{-1mm}
-~{\sigma\beta_Bs\over \sigma\beta_Bs+k_\perp^2}\ticalf_i(\beta_B, x_\perp)\calf_j(\beta_B, y_\perp)\Big]\Big\}|p\rangle
~+~O(\alpha_s^2),
\nonumber
\end{eqnarray}
where we use Schwinger's notations
\begin{equation}
(x_\perp|f(p_\perp)|y_\perp)~\equiv~ \int\! \dhd^2p_\perp~e^{i(p,x-y)_\perp}f(p), ~~~~~(x_\perp|p_\perp)~=~e^{i(p,x)_\perp}.
\label{schwingerperp}
\end{equation}
Operator $U=[\infty, -\infty]$ is an infinite Wilson line.

The equation looks rather untransparent. However, we can show that in different kinematic limits this equation significantly simplifies and reproduces many well-known results. I start with the moderate-x limit, when $\beta_B=x_B\sim1$ and the transverse momentum of the "detected" parton $q^2_\perp\sim(x-y)^{-2}_\perp\sim s$ is of the order of the hard scale of the problem. Then I move to the small-x, when both Bjorken $\beta_B\equiv x_B\sim\frac{(x-y)^{-2}}{s}$ and $q^2_\perp\sim (x-y)^{-2}_\perp\ll s$ have small value. After that I discuss an intermediate limit, when $\beta_B\equiv x_B\sim1$ is large, but $q^2_\perp\sim(x-y)^{-2}_\perp\ll s$ is still small. This limit reveals the Sudakov double log evolution. Finally I write a linearized version of the evolution equation (\ref{masterdis}) and show that it combines DGLAP \cite{dglap} and BFKL \cite{BFKL} logarithms. I this this equation is very promising for future phenomenological studies at EIC.

\section{Moderate-x limit}

In this section we consider the moderate-x (DGLAP) limit of the evolution equation, when $\beta_B=x_B\sim 1$ and $k^2_\perp\sim (x-y)^{-2}_\perp\sim s$. In this case the main role plays a kinematic condition $\theta\big(1-\beta_B-{k_\perp^2\over\sigma s}\big)$ in the real gluon emission part. The origin of this kinematic restriction is obvious: if $p$ is a momentum of the parton before the emission, the sum of the longitudinal momentum fraction of the detected parton $\beta_B p$ and the emitted gluon $\frac{k^2_\perp}{\sigma s}p$ can not exceed its value.

A presence of such kinematic condition means a strong correlation between the rapidity of the emitted particle and its transverse momentum. Indeed, according to this condition $k^2_\perp<\sigma s(1-\beta_B)$. If we follow the evolution and go from some value of rapidity cut-off $\sigma$ to smaller values $\sigma'\ll\sigma$, we immediately get $k'^2_\perp\ll k^2_\perp$.

In the rapidity factorization approach we separate all fields into the "quantum" fields with $\alpha>\sigma$ and "classical" fields with $\alpha<\sigma$. The presence of the kinematic restriction means that there is ordering of the transverse momentum $l^2_\perp\ll k^2_\perp$ as well, where $k_\perp$ ($l_\perp$) is a typical transverse momentum of the quantum (classical) fields. That means that in the moderate-x limit we are effectively in the light-cone limit: one can perform expansion of the evolution kernel onto the light-cone direction and neglect contribution suppressed by the ration $l_\perp/k_\perp\ll 1$. One can check that the leading contribution in this case comes from contributions linear in the background field strength tensor $F_{\bullet i}$ and the general equation reduces to
\begin{eqnarray}
&&\hspace{-1mm}
{d\over d\ln\sigma}\langle p|\ticalf_i^a(\beta_B,x_\perp)\calf_j^a(\beta_B,y_\perp)|p\rangle~
\label{3.21}\\
&&\hspace{-1mm}
=~{g^2N_c\over\pi}\!\int\!\dhd^2k_\perp~\Big\{e^{i(k,x-y)_\perp}
\langle p|\ticalf_k^a\big(\beta_B+{k_\perp^2\over\sigma s},x_\perp\big)\calf_l^a\big(\beta_B+{k_\perp^2\over\sigma s},y_\perp\big)
|p\rangle
\nonumber\\
&&\hspace{-1mm}
\times~\Big[{\delta_i^k\delta_j^l\over k_\perp^2}-{2\delta_i^k\delta_j^l\over\sigma\beta_Bs+k_\perp^2}
+~{k_\perp^2\delta_i^k\delta_j^l+\delta_j^kk_ik^l+\delta_i^lk_jk^k-\delta_j^lk_ik^k-\delta_i^kk_jk^l-g^{kl}k_ik_j-g_{ij}k^kk^l
\over (\sigma\beta_Bs+k_\perp^2)^2}
\nonumber\\
&&\hspace{2mm}
+~k_\perp^2{2g_{ij}k^kk^l+\delta_i^kk_jk^l+\delta_j^lk_ik^k-\delta_j^kk_ik^l-\delta_i^lk_jk^k\over (\sigma\beta_Bs+k_\perp^2)^3}
-{k_\perp^4g_{ij}k^kk^l\over  (\sigma\beta_Bs+k_\perp^2)^4}\Big]
\theta\big(1-\beta_B-{k_\perp^2\over \sigma s}\big)
\nonumber\\
&&\hspace{58mm}
-~ {\sigma\beta_Bs\over k_\perp^2(\sigma\beta_Bs+k_\perp^2)} 
\langle p|\ticalf_i^a(\beta_B,x_\perp)\calf_j^a(\beta_B,y_\perp)|p\rangle\Big\}.
\nonumber
\end{eqnarray}
One can consider a special case of unpolarized scattering and rewrite evolution in terms of two gluon distribution functions $\mathcal{D}(\beta_B, z_\perp, \ln\sigma)$ and $\mathcal{H}(\beta_B, z_\perp, \ln\sigma)$ \cite{muldrod}:
\begin{eqnarray}
&&\hspace{-1mm}
\langle p|\ticalf_i^a(\beta_B,z_\perp)\calf_j^a(\beta_B,0_\perp)|p+\xi p_2\rangle^\eta~
\nonumber\\
&&\hspace{2mm}
=~
 2\pi^2\delta(\xi) \beta_Bg^2\Big[-g_{ij}\cald(\beta_B,z_\perp,\eta)
-{4\over m^2}(2z_iz_j+g_{ij}z_\perp^2)\calh''(\beta_B,z_\perp,\eta)\Big],
\label{decomp}
\end{eqnarray}
where $\calh''(\beta_B,z_\perp,\eta)\equiv\big({\partial\over\partial z^2}\big)^2\calh(\beta_B,z_\perp,\eta)$. We substitute this decomposition into (2.1), which yields a system of two equations:
\begin{eqnarray}
&&\hspace{-1mm}
{d\over d\eta}\alpha_s\cald(\beta_B,z_\perp,\eta)
\label{liconefinal}\\
&&\hspace{-1mm}
=~ {\alpha_sN_c\over \pi}\!\int_{\beta_B}^1\!{dz'\over z'}\Big\{
J_0\Big(|z_\perp|\sqrt{\sigma s\beta_B{1-z'\over z'}}\Big)
\Big[\big({1\over 1-z'}\big)_+ +{1\over z'}-2+z'(1-z')\Big]\alpha_s\cald\big({\beta_B\over z'},z_\perp,\eta\big)   
\nonumber\\
&&\hspace{-1mm}
+~  
{4\over m^2}(1-z')z'
z_\perp^2J_2\Big(|z_\perp|\sqrt{\sigma s\beta_B{1-z'\over z'}}\Big)
\alpha_s\calh''({\beta_B\over z'},z_\perp,\eta)\Big\},
\nonumber\\
&&\hspace{-1mm}
{d\over d\eta}\alpha_s\calh''(\beta_B,z_\perp,\eta)
\nonumber\\
&&\hspace{-1mm}
=~ {\alpha_sN_c\over \pi}\!\int_{\beta_B}^1\!{dz'\over z'}\Big\{
J_0\Big(|z_\perp|\sqrt{\sigma s\beta_B{1-z'\over z'}}\Big)
\Big[\big({1\over 1-z'}\big)_+ -1\Big]\alpha_s\calh''\big({\beta_B\over z'},z_\perp,\eta\big)
\nonumber\\
&&\hspace{-1mm}
+~~{m^2\over 4z_\perp^2}{1-z'\over z'}J_2\Big(|z_\perp|\sqrt{\sigma s\beta_B{1-z'\over z'}}\Big)
\alpha_s\cald\big({\beta_B\over z'},z_\perp,\eta\big)\Big\},
\nonumber
\end{eqnarray}
where $z'\equiv {\beta_B\over\beta+\beta_B}$. We see that there is an entanglement in evolution between two unpolarized gluon distribution functions. It is important that in the collinear limit $x_\perp=y_\perp$ the first equation reduces to the well-known DGLAP equation \cite{dglap}:
\begin{equation}
\hspace{-1mm}
{d\over d\eta}\alpha_s\cald(\beta_B,0_\perp,\eta)~=~{\alpha_s\over\pi}N_c
\!\int_{\beta_B}^1\! {dz'\over z'}~
\Big[\big({1\over 1-z'}\big)_+  +{1\over z'}- 2+ z'(1-z')\Big]
\alpha_s\cald\big({\beta_B\over z'},0_\perp,\eta\big).
\label{DGLAP}
\end{equation}
As a result we see that the general equation (\ref{masterdis}) contains the moderate-x dynamics.

\section{Small-x limit\label{smallxsec}}
In this limit we consider evolution equation (\ref{masterdis}) at small values of $\beta_B\ll 1$. In the small-x limit all transverse momenta are of the same order. We also assume that characteristic transverse momentum imposed by the integral over $k^2_\perp$ in the real emission part is of the order of $(x-y)^{-2}_\perp\ll s$. As a result in the range of rapidity evolution ${(x-y)_\perp^{-2}\over s}\ll\sigma\ll {(x-y)_\perp^{-2}\over \beta_Bs}$ we get $\frac{p^2_\perp}{\sigma s}\ll1$ (the upper bound imposes that there is no cancellation of the non-linear part). In this case the following relation holds
\begin{eqnarray}
\calf_i\big(\beta_B+{p_\perp^2\over\sigma s}\big)\simeq\calf_i(\beta_B)\simeq i\partial_iUU^\dagger.
\label{smallxestim}
\end{eqnarray}
A careful examination of the general evolution equation (\ref{masterdis}) gives the following small-x form:
\begin{eqnarray}
&&\hspace{-1mm}
{d\over d\ln\sigma}\tilU^a_i(x_\perp)U^a_j(y_\perp)~
=~-4\alpha_s{\rm Tr}\Big\{\big(x_\perp\big|
\tilU p_i\tilU^\dagger\big(\tilU{p^k\over p_\perp^2}\tilU^\dagger 
-{p^k\over p_\perp^2}\big)\big(U{p_k\over p_\perp^2}U^\dagger 
-{p_k\over p_\perp^2}\big)Up_j U^\dagger|y_\perp)
\nonumber\\
&&\hspace{37mm}
-~\Big[
(x_\perp|\tilU{p_ip^k\over p_\perp^2}\tilU^\dagger{p_k\over p_\perp^2}|x_\perp)
-\half(x_\perp|{1\over p_\perp^2}|x_\perp)\tilU_i(x_\perp)\Big]
U_j(y_\perp)
\nonumber\\
&&\hspace{44mm}
-~\tilU_i(x_\perp)\Big[
(y_\perp|{p^k\over p_\perp^2}U{p_jp_k\over p_\perp^2}U^\dagger|y_\perp)
-\half(y_\perp|{1\over p_\perp^2}|y_\perp)U_j
(y_\perp)\Big]\Big\}.
\label{nonlinfirst}
\end{eqnarray}
Note that the TMD operator $\ticalf_i^a(\beta_B,x_\perp)\calf_j^a(\beta_B,y_\perp)$ reduces to the Weizsacker-Williams (WW) distribution $\tilU^a_i(x_\perp)U^a_j(y_\perp)$, which is independent of $\beta_B$.

At this point it is useful to utilize the following formula $(x_\perp|\frac{p_i}{p^2_\perp}|y_\perp)=\int\dhd^2p_\perp\frac{p_i}{p^2_\perp}e^{ip_\perp(x-y)_\perp}=\frac{1}{2\pi i}\frac{(x-y)_i}{(x-y)^2}$ and apply it to (\ref{nonlinfirst}), which after fairly long calculations yields
\begin{eqnarray}
&&\hspace{-1mm} 
{d\over d\eta}\tilU^a_i(z_1) U^a_j(z_2)
\label{nonlinsec}\\
&&\hspace{-1mm}
=~-{g^2\over 8\pi^3}{\rm Tr}\big\{
(-i\partial^{z_1}_i+\tilU^{z_1}_i)\big[\!\int\! d^2z_3(\tilU_{z_1}\tilU^\dagger_{z_3}-1)
{z_{12}^2\over z_{13}^2z_{23}^2}(U_{z_3}U^\dagger_{z_2}-1)\big]
(i\stackrel{\leftarrow}{\partial^{z_2}_j}+U^{z_2}_j)\big\}.
\nonumber 
\end{eqnarray}
It would be interesting to compare this equation with the small-x TMD evolution in Ref. \cite{domumuxi} obtained from the JIMWLK equation \cite{JIMWLK}.
We see that equation (\ref{masterdis}) describes the non-linear behavior of gluon TMDs typical for small-x, cf. \cite{npb96, yura}.

\section{Sudakov limit}
In the Sudakov limit we consider a case of moderate $x_B\equiv\beta_B \sim 1$, but relatively small momentum of the detected parton $q^2_\perp\sim (x-y)^{-2}_\perp\sim{\rm few \ GeV^2}\ll s$. First of all, we expect the integral over $k^2_\perp$ in the real emission part to fall into $k^2_\perp\sim q^2_\perp\sim (x-y)^{-2}_\perp$. The only thing that could prevent this is the kinematic restriction $\theta\big(1-\beta_B-{k_\perp^2\over\sigma s}\big)$, which potentially could cut out this region. However, if we keep the evolution in the range of $\frac{(x-y)^{-2}_\perp}{s}\ll\sigma\ll1$, we see that for $k^2_\perp\sim(x-y)^{-2}_\perp$ one can estimate $\frac{k^2_\perp}{\sigma s}\ll 1$ and neglect the $\theta$-function. That in turn means that typical intermediate momentum is also restricted to $p^2_\perp\sim (x-y)^{-2}_\perp$. As a result one can neglect any combination of $\frac{p^2_\perp}{\sigma \beta_B s}\sim\frac{k^2_\perp}{\sigma \beta_Bs}\ll 1$ in the real emission part (it is important here that $\beta_B\sim1$). We also take into account that a typical transverse momentum of the background fields $l_\perp\lesssim p_\perp$, so the structures with transverse derivative, i.e. $p_l\partial^lU/\sigma \beta_Bs\sim p_l\mathcal{F}^l/\sigma \beta_Bs\ll1$, can be omitted as well. All together this significantly simplifies the real emission part - almost all terms can be neglected. The only term that survives is the one without the suppression factor $1/\sigma\beta_B s$, i.e. (see the last but one line of the evolution equation)
\begin{eqnarray}
4\alpha_sN_c\!\int\!{\dhd^2 p_\perp\over p_\perp^2}e^{i(p,x-y)_\perp} 
\langle p|\tilcaf^a_i\big(\beta_B+{p_\perp^2\over\sigma s},x_\perp\big)
\calf^a_j\big(\beta_B+{p_\perp^2\over\sigma s},y_\perp\big)|p\rangle.
\nonumber
\end{eqnarray}

We see that non-linear contribution from the real gluon emission can be neglected. The same is true for the virtual part, at least with our accuracy. Indeed, let's look at one of the non-linear terms
\begin{eqnarray}
&&(y_\perp|-{p^m\over p_\perp^2}\calf_k(\beta_B)(i\!\stackrel{\leftarrow}{\partial}_l+U_l)(2\delta_m^k\delta_j^l-g_{jm}g^{kl})
U{1\over \sigma\beta_Bs+p_\perp^2}U^\dagger|y_\perp)
\end{eqnarray}
and expand operator $\calo~\equiv~\calf_k(\beta_B)(i\!\stackrel{\leftarrow}{\partial}_l+U_l)(2\delta_m^k\delta_j^l-g_{jm}g^{kl})U$ around $y_\perp$ direction: $\calo_{z_\perp}~=~\calo_{y_\perp}+(y-z)^i\partial_i\calo_{y_\perp}+\dots$. One can explicitly integrate over $p_\perp$ and find that
\begin{eqnarray}
&&\hspace{-1mm} 
(y_\perp|{p_m\over p_\perp^2}\calo{1\over \sigma\beta_Bs+p_\perp^2}|y_\perp)~
=~
\calo_y(y_\perp|{p_m\over p_\perp^2(\sigma\beta_Bs+p_\perp^2)}|y_\perp)
+{i\partial_m\calo_y\over 4\pi\sigma\beta_Bs}+\dots .
\nonumber
\end{eqnarray}
The first term is obviously zero and the second one is negligibly small. As a result we see that non-linear contribution doesn't survive in the Sudakov limit. The general evolution equation takes a very simple form:
\begin{eqnarray}
&&\hspace{-1mm}
{d\over d\ln\sigma}\langle p|\tilcaf^a_i(\beta_B, x_\perp) \calf_j^a(\beta_B, y_\perp)|p\rangle~
\label{sudak}\\
&&\hspace{-1mm}
=~4\alpha_sN_c\!\int\!{\dhd^2 p_\perp\over p_\perp^2}\Big[e^{i(p,x-y)_\perp} 
\langle p|\tilcaf^a_i\big(\beta_B+{p_\perp^2\over\sigma s},x_\perp\big)
\calf^a_j\big(\beta_B+{p_\perp^2\over\sigma s},y_\perp\big)|p\rangle
\nonumber\\
&&\hspace{44mm}
-~{\sigma \beta_Bs\over\sigma \beta_Bs+p_\perp^2}\langle p|\tilcaf^a_i(\beta_B,x_\perp)\calf^a_j(\beta_B,y_\perp)|p\rangle
\Big].
\nonumber
\end{eqnarray}
We see that the real contribution comes from the region of $p_\perp\sim(x-y)^{-2}_\perp$, while there is no formal restriction for transverse momentum in the virtual part. However, it is effectively suppressed in the UV region by the $1/(\sigma\beta_B s+p^2_\perp)$ factor, so we can introduce a cut-off $p^2_\perp<\sigma\beta_B s$ in the virtual term. As a result, there will be a combination of virtual and real parts for small $p^2_\perp\sim (x-y)^{-2}_\perp$ (with cancellation of the IF divergence for $p^2_\perp\to 0$) and a region of large $(x-y)^{-2}_\perp\ll p^2_\perp \ll \sigma\beta_B s\sim\sigma s$, which provides only virtual emission. The first region gives us a single-log evolution, while the second one introduces the Sudakov double-log. For the latter one we have
\begin{eqnarray}
&&\hspace{-1mm} 
{d\over d\ln\sigma}\langle p|\tilcaf^a_i(\beta_B, x_\perp) \calf_j^a(\beta_B, y_\perp)|p\rangle
~\simeq~-{\alpha_sN_c\over \pi}\langle p|\tilcaf^a_i(\beta_B, x_\perp) \calf_j^a(\beta_B, y_\perp)|p\rangle
\ln\sigma s z_\perp^2
\label{7.9}
\end{eqnarray}
with a double-log solution for the distribution function:
\begin{eqnarray}
&&\hspace{-1mm} 
\cald(x_B,k_\perp,\ln\sigma)
~\sim~\exp\big\{-{\alpha_sN_c\over 2\pi}\ln^2{\sigma s\over k_\perp^2} \big\}\cald(x_B,k_\perp,\ln{k_\perp^2\over s}).
\label{sudakov}
\end{eqnarray}
It is worth noting that the coefficient in front of $\ln^2{\sigma s\over k_\perp^2}$ is determined by the cusp anomalous dimension \cite{cusp} of two light-like Wilson lines going from point $y$ to $\infty p_1$ and $\infty p_2$ directions (with our cutoff $\alpha<\sigma$).

\section{Linearization}
 It is interesting to obtain a linearized version of the evolution equation (\ref{masterdis}). This implies that we keep only one gluon filed at each side from the cut and neglect all multi-gluon interactions, which corresponds to the Wilson line factors $U$ and $\tilde{U}$. We expect that the simplified equation will reproduce explicitly linear evolutions of DGLAP at moderate x and BFKL at small x. This equation will provide a smooth transition between two limits.
 
The procedure is pretty straightforward: we keep operators $\ticalf^k\big(\beta_B+{k_\perp^2\over\sigma s}, x_\perp\big)$ and
$\calf^l\big(\beta_B+{k_\perp^2\over\sigma s}, y_\perp\big)$, which correspond to two single gluons coming from the background field, and drop all other fields from the equation. This yields 
\begin{eqnarray}
&&\hspace{-3mm}
{d\over d\ln\sigma}\langle p|\tilcaf_i^a(\beta_B, p_\perp) \calf_j^a(\beta_B, p'_\perp)|p\rangle~
\label{masterlin}\\
&&\hspace{-3mm}
=~-\alpha_sN_c\!\int\!\dhd^2k_\perp
\Big\{\theta\big(1-\beta_B-{k_\perp^2\over\sigma s}\big)
\Big[\Big({(p+k)_k\over\sigma\beta_Bs+p_\perp^2}
{\sigma \beta_Bsg_{\mu i}-2k^\perp_{\mu}k_i\over\sigma \beta_Bs+k_\perp^2}
-~2{k^\perp_\mu g_{ik}+p_ig_{\mu k} \over \sigma\beta_Bs+p_\perp^2}\Big)
\nonumber\\
&&\hspace{55mm}
\times~
\Big({\sigma \beta_Bs\delta^\mu_j-2k_\perp^{\mu}k_j\over\sigma\beta_Bs+k_\perp^2}{(p'+k)_l\over\sigma \beta_Bs+{p'}_\perp^2}
-2{k_\perp^\mu g_{jl}+\delta_l^\mu p'_j\over \sigma\beta_Bs+{p'}_\perp^2}\Big)
\nonumber\\
&&\hspace{5mm}
+~2g_{ik}\Big({k_j\over k_\perp^2}{\sigma \beta_Bs+2k_\perp^2\over\sigma\beta_Bs+k_\perp^2}
{(p'+k)_l\over\sigma \beta_Bs+{p'}_\perp^2}
+~{2g_{jl}\over \sigma\beta_Bs+{p'}_\perp^2}-{2p'_jk_l\over k_\perp^2(\sigma\beta_Bs+{p'}_\perp^2)}
\Big)
\nonumber\\
&&\hspace{5mm}
+~2g_{lj}\Big({(p+k)_k\over\sigma\beta_Bs+p_\perp^2}
{k_i\over k_\perp^2}
{\sigma \beta_Bs+2k_\perp^2\over\sigma \beta_Bs+k_\perp^2}
+{2g_{ik}\over \sigma\beta_Bs+p_\perp^2}
-~{2p_ik_k \over k_\perp^2(\sigma\beta_Bs+p_\perp^2)}\Big)\Big]
\nonumber\\
&&\hspace{55mm}
\times~
\langle p|\ticalf^k\big(\beta_B+{k_\perp^2\over\sigma s}, p_\perp-k_\perp\big)
\calf^l\big(\beta_B+{k_\perp^2\over\sigma s}, p'_\perp-k_\perp\big)|p\rangle
\nonumber\\
&&\hspace{5mm}
+~{2\over k_\perp^2}\Big[
{(2k^lp'_j-k_j{p'}^l)\delta_i^k\over \sigma\beta_Bs+(p'+k)_\perp^2}
+~{(2p_ik^k-k_ip^k)\delta_j^l\over \sigma\beta_Bs+(p+k)_\perp^2}\Big]
\langle p|\ticalf_k^a(\beta_B, p_\perp)\calf_l^a(\beta_B, p'_\perp)|p\rangle
\nonumber\\
&&\hspace{5mm}
-~{4\over k_\perp^2}\langle p|
\Big[\theta\big(1-\beta_B-{k_\perp^2\over\sigma s}\big)
\ticalf_i\big(\beta_B+{k_\perp^2\over\sigma s}, p_\perp-k_\perp\big)
\calf_j\big(\beta_B+{k_\perp^2\over\sigma s}, p'_\perp-k_\perp\big)
\nonumber\\
&&\hspace{55mm}
-~{\sigma\beta_Bs\over \sigma\beta_Bs+k_\perp^2}\ticalf_i^a(\beta_B, p_\perp)\calf_j^a(\beta_B, p'_\perp)\Big]|p\rangle
\Big\},
\nonumber
\end{eqnarray}
where we performed Fourier transformation to the momentum space using definitions (\ref{kalf}) and (\ref{tilkalf}). For the terms with derivatives we've used
\begin{eqnarray}
&&2\tilde{\mathcal{F}}_i(\beta_B, p_\perp)\int d^2y_\perp e^{-ip'_\perp y_\perp}
(y_\perp|-\frac{p^m}{p^2_\perp} i\partial_l\mathcal{F}_k(\beta_B)\frac{2\delta^k_m\delta^l_j-g_{jm}g^{kl}}{\sigma\beta_B s+p^2_\perp} |y_\perp)
\\
&&=-2\tilde{\mathcal{F}}_i(\beta_B, l_\perp)\int \dhd^2k_{\perp}
\frac{(k+p')^m}{(k+p')^2_\perp}p'_l\mathcal{F}_k(\beta_B, p'_\perp)\frac{2\delta^k_m\delta^l_j-g_{jm}g^{kl}}{\sigma\beta_B s+k^2_{\perp}}
\nonumber
\end{eqnarray}
and a similar expression for the $i\partial_k \tilde{\mathcal{F}}_l(\beta_B)$ operator. It is also useful to note that for arbitrary operator $A$ defined in the coordinate space as $A|x)=A(x)|x)$ we have a Fourier transform $(p_1|A|p_2)=A(p_1-p_2)$ and $(p_1|\tilde{A}|p_2)=\tilde{A}(-p_1+p_2)$ for the complex conjugated side.

By means of decomposition (\ref{decomp}) we write the matrix element as
\begin{eqnarray}
&&\langle p| \tilde{\mathcal{F}}^a_i(\beta_B, p_\perp)\mathcal{F}^{a}_j(\beta_B, p'_\perp)|p+\eta p_2\rangle= (2\pi)^2\delta^{(2)}(p_\perp-p'_\perp)2\pi^2\delta(\xi)\beta_Bg^2\mathcal{R}_{ij}(\beta_B, p_\perp),
\nonumber
\end{eqnarray}
where
\begin{eqnarray}
&&\mathcal{R}_{ij}(\beta_B, p_\perp)=-g_{ij}\mathcal{D}(\beta_B,p_\perp)+(\frac{2p_ip_j}{m^2}+g_{ij}\frac{p^2_\perp}{m^2})\mathcal{H}(\beta_B, p_\perp).
\end{eqnarray}
In terms of $\mathcal{R}_{ij}(\beta_B, p_\perp)$ the linearized version of the evolution equation (\ref{masterdis}) takes its final form
\begin{eqnarray}
&&\hspace{-1mm}
{d\over d\ln\sigma}\calr_{ij}(\beta_B,p_\perp;\eta)
\label{linear}\\
&&\hspace{-1mm}
=~-\alpha_sN_c\!\int\!\dhd^2k_\perp\Big\{
\Big[\Big({(2p-k)_k\over\sigma\beta_Bs+p_\perp^2}
{\sigma \beta_Bsg_{\mu i}-2(p-k)^\perp_{\mu}(p-k)_i\over\sigma \beta_Bs+(p-k)_\perp^2}
-~2{(p-k)^\perp_\mu g_{ik}+p_ig_{\mu k} \over \sigma\beta_Bs+p_\perp^2}\Big)
\nonumber\\
&&\hspace{22mm}
\times~
\Big({\sigma \beta_Bs\delta^\mu_j-2(p-k)_\perp^{\mu}(p-k)_j\over\sigma\beta_Bs+(p-k)_\perp^2}
{(2p-k)_l\over\sigma \beta_Bs+p_\perp^2}
-2{(p-k)_\perp^\mu g_{jl}+\delta_l^\mu p_j\over \sigma\beta_Bs+p_\perp^2}\Big)
\nonumber\\
&&\hspace{-1mm}
+~2g_{ik}
\Big({(p-k)_j(2p-k)_l-2p_j(p-k)_l\over (p-k)_\perp^2(\sigma\beta_Bs+p_\perp^2)}
+{(p-k)_j(2p-k)_l\over (\sigma\beta_Bs+(p-k)_\perp^2)(\sigma\beta_Bs+p_\perp^2)}
+~{2g_{jl}\over \sigma\beta_Bs+p_\perp^2}\Big)
\nonumber\\
&&\hspace{-1mm}
+~2g_{lj}\Big(
{(p-k)_i(2p-k)_k-2p_i(p-k)_k\over (p-k)_\perp^2(\sigma\beta_Bs+p_\perp^2)}
+{(p-k)_i(2p-k)_k\over (\sigma\beta_Bs+(p-k)_\perp^2)(\sigma\beta_Bs+p_\perp^2)}
+{2g_{ik}\over \sigma\beta_Bs+p_\perp^2}\Big)
\Big]
\nonumber\\
&&\hspace{55mm}
\times~
\theta\big(1-\beta_B-{(p-k)_\perp^2\over\sigma s}\big)\calr^{kl}\big(\beta_B+{(p-k)_\perp^2\over\sigma s},k_\perp\big)
\nonumber\\
&&\hspace{-1mm}
+~2
{\delta_i^k(k_jp^l-2k^lp_j)+\delta^l_j(k_ip^k -2p_ik^k)\over k_\perp^2[\sigma\beta_Bs+(p-k)_\perp^2]}
\calr_{kl}(\beta_B,p_\perp;\eta)
\nonumber\\
&&\hspace{-1mm}
-~
4\Big[{\theta\big(1-\beta_B-{(p-k)_\perp^2\over\sigma s}\big)\over (p-k)_\perp^2}
\calr_{ij}\big(\beta_B+{(p-k)_\perp^2\over \sigma s},k_\perp;\eta\big)
-~{\sigma\beta_Bs\over k_\perp^2(\sigma\beta_Bs+k_\perp^2)}\calr_{ij}(\beta_B,p_\perp;\eta)\Big]\Big\}.
\nonumber
\end{eqnarray}
One should note that this equation describes a forward scattering, which implies that there is no transition of the transverse momentum along the gluon ladder, i.e. $p_\perp=p'_\perp$. For study of the non-forward case one should stay with a more general evolution equation (\ref{masterlin}).

At this point we should demonstrate that equation (\ref{linear}) contains both DGLAP and BFKL dynamics. Let's start with the small-x limit. As was showen in Section \ref{smallxsec}, we can set $\beta_B=0$ and neglect $\frac{k^2_\perp}{\sigma s}\sim 0$. According to (\ref{smallxestim}) the matrix element reduces to the WW distribution:
\begin{eqnarray}
&&\calr_{ij}(0,k_\perp)~\sim ~\int\! d^2xe^{i(k,x)_\perp}\langle p|{\rm tr}\{\tilU_i(x)U_j(0)\}|p\rangle.
\end{eqnarray}
It is well known (cf. Ref. \cite{nlobk}) that in the leading-order BFKL approximation
\begin{eqnarray}
&&\hspace{-1mm}
\langle p|{\rm tr}\{\tilU_i(x)U_j(y)\}|p\rangle~
\label{lip1}\\
&&\hspace{-1mm}
=~{\alpha_s\over 4\pi^2}\!\int\!{d^2 q_\perp\over q_\perp^2}
q_iq_je^{i(q,x)_\perp-i(q,y)_\perp}\!\int\!{d^2 q'_\perp\over {q'}_\perp^2}\Phi_T(q')
\!\int_{a-i\infty}^{a+i\infty}\!{d\omega\over 2\pi i}\Big({s\over qq'}\Big)^\omega
G_\omega(q,q').
\nonumber
\end{eqnarray}
Here $\Phi_T(q')$ is the target impact factor 
and  $G_\omega(q,q')$ is the partial wave of the forward reggeized gluon scattering amplitude satisfying the equation
\begin{equation}
\omega G_\omega(q,q')=\delta^{(2)}(q-q')+\int\! d^2p K_{\rm BFKL}(q,p)G_\omega(p,q') 
\label{wgw}
\end{equation}
with the forward BFKL kernel 
$$
K_{\rm BFKL}(q,p)~=~{\alpha_sN_c\over\pi^2}
\Big[{1\over (q-p)_\perp^2}-\half\delta(q_\perp-p_\perp)\!\int\!dp'_\perp{q_\perp^2\over {p'}_\perp^2(q-p')_\perp^2}\Big].
$$
Thus, to get the forward BFKL evolution equation we substitute
\begin{equation}
\calr_{ij}(0,q_\perp;\ln\sigma)~=~q_iq_jR(q_\perp;\ln\sigma)
~=~{\alpha_sq_iq_j\over 2\pi^2q_\perp^2}\!\int\!
{d^2 q'\over {q'}^2}\Phi_T(q')
\!\int_{a-i\infty}^{a+i\infty}\!{d\omega\over 2\pi i}\Big({\sigma s\over qq'}\Big)^\omega
G_\omega(p,q')
\label{R}
\end{equation}
into Eq. (\ref{linear}) and after some algebra get
\begin{eqnarray}
&&\hspace{-1mm}
{d\over d\ln\sigma}R(p_\perp;\ln\sigma)
~=~2\alpha_sN_c\!\int\!\dhd^2k_\perp
\Big[{2k_\perp^2\over p_\perp^2(p-k)_\perp^2}
R(k_\perp;\ln\sigma)
-~{p_\perp^2\over k_\perp^2(p-k)_\perp^2}R(p_\perp;\ln\sigma)\Big],
\nonumber
\end{eqnarray}
which is the forward scattering BFKL equation \cite{BFKL}. We have also checked that Eq. (\ref{masterlin}) at $p_\perp\neq p'_\perp$ reduces to the non-forward BFKL equation in the low-$x$ limit: one just need to substitute
\begin{eqnarray}
\langle p| \tilde{\mathcal{F}}^a_i(\beta_B, p_\perp)\mathcal{F}^{a}_j(\beta_B, p_\perp-q_\perp)|p+\eta p_2\rangle\sim p_i(p-q)_jR(p_\perp, q_\perp).
\end{eqnarray}

The last thing we need to check is how the evolution of 
\begin{equation}
\cald(\beta_B,\ln\sigma)~=~-\half\int\! \dhd^2 p_\perp \calr_i^{~i}(\beta_B,p_\perp;\ln\sigma)
\label{rii}
\end{equation}
reduces to the DGLAP equation.  
As we discussed above, in the light-cone limit one can neglect $k_\perp$ in comparison to $p_\perp$. 
Indeed, the integral over $p_\perp$ converges at $p_\perp^2\sim\sigma\beta_Bs$. On the other hand, extra $k_ik_j$ in the integral 
over $k_\perp$ leads to the operators of higher collinear twist, for example
\begin{eqnarray}
&&\hspace{0mm} 
\int\! d^2k_\perp ~k_ik_j~R_n^{~n}(\beta_B,k_\perp;\ln\sigma)
~\sim~\langle p|\partial_k\tilcaf^a_n(\beta_B, 0_\perp)\partial_j\calf^{an}(\beta_B, 0_\perp)|p\rangle^{\eta=\ln\sigma}
\nonumber\\
&&\hspace{0mm}
\sim~m^2g_{ij}\langle p|\tilcaf^a_n(\beta_B, 0_\perp)\calf^{a n}(\beta_B, 0_\perp)|p\rangle^{\ln\sigma}~\sim~m^2\cald(\beta_B,\ln\sigma)
\end{eqnarray}
(where $m$ is the mass of the target), so ${k_\perp^2\over p_\perp^2}\sim{k_\perp^2\over\sigma\beta_Bs}\sim {m^2\over\sigma s}\ll 1$.

Neglecting $k_\perp$ in comparison to $p_\perp$ and integrating over angles one obtains
\begin{eqnarray}
&&\hspace{-1mm}
{d\over d\ln\sigma}\!\int\! d^2p_\perp~\calr_i^{~i}(\beta_B,p_\perp;\ln\sigma)
\\
&&\hspace{-1mm}
=~{\alpha_sN_c\over \pi^2}\!\int\! d^2p_\perp\Big[{1\over p_\perp^2}
-{2\over\sigma\beta_Bs+p_\perp^2}+{3p_\perp^2\over(\sigma\beta_Bs+p_\perp^2)^2}
-{2p_\perp^4\over(\sigma\beta_Bs+p_\perp^2)^3}+{p_\perp^6\over(\sigma\beta_Bs+p_\perp^2)^4}\Big]
\nonumber\\
&&\hspace{-1mm}
\times~\!\int\! d^2k_\perp
\calr_i^{~i}\big(\beta_B+{p_\perp^2\over\sigma s},k_\perp;\ln\sigma\big)
-\!\int\! d^2k_\perp{\sigma\beta_Bs\over k_\perp^2(\sigma\beta_Bs+k_\perp^2)}\!\int\! d^2p_\perp\calr_i^{~i}\big(\beta_B,p_\perp;\ln\sigma\big)
\nonumber
\end{eqnarray}
One can introduce new variables $z'=\frac{\beta_B}{\beta_B+\beta}$, $\beta\equiv\frac{k^2}{\sigma s}$ and show that this equation reduces to the DGLAP equation (\ref{DGLAP}).

 It would be interesting to compare Eq. (\ref{masterlin}) to CCFM equation \cite{ccfm} which also
addresses the question of interplay of BFKL and DGLAP logarithms.

\section{Conclusions}
We show how the rapidity evolution equation (\ref{masterdis}) looks in different kinematic limits. We find that it significantly simplifies and coincides with many previously known results. That corroborates that the equation provides a correct answer in the whole range of Bjorken $x_B$ and the whole range of transverse momentum $k_\perp$.

In particular, we look at three limits. It is worth to say that they correspond to different steps of evolution. Indeed, at initial stage the DGLAP evolution is important. In this case $x_B\equiv\beta_B\sim1$ and $k^2_\perp\sim s$. We show that in this case evolution in rapidity is correlated with the value of $k_\perp$ through a kinematic condition $\theta\big(1-\beta_B-{k_\perp^2\over\sigma s}\big)$. If $\sigma$ is not too small, i.e. $\sigma\lesssim 1$, this condition brings a strict ordering of the transverse momenta. All non-linear terms can be neglected and we get a system of two linear evolution equations (\ref{liconefinal}) of the DGLAP-type.

However, if we go to smaller $\sigma$ this $\theta$-function effectively cut-off the DGLAP evolution. At this stage we take smaller values of $k^2_\perp\sim(x-y)^{-2}_\perp\ll s$, but keep moderate $x_B\equiv\beta_B\sim1$. In the region $\frac{(x-y)^{-2}}{\beta_Bs}\ll\sigma\ll 1$ we can neglect the kinematic condition and the non-linear terms as well. This leads to a linear evolution equation of the Sudakov type (\ref{sudak}).

If we continue evolution to even smaller values ${(x-y)^{-2}_\perp\over s}\ll\sigma\ll{(x-y)^{-2}_\perp\over\beta_Bs}$, the only evolution that survives is the small-x evolution with $x_B\equiv\beta_B\ll1$ and $k^2_\perp\sim (x-y)^{-2}_\perp\ll s$. We can omit the $\theta$-function, but can no longer neglect the non-linear part. We talk about the shock-wave picture of interaction and get a non-linear equation (\ref{nonlinsec}) of the BK-type.

As a result we see how the general evolution equation (\ref{masterdis}) describes an interplay between different logarithms and gives a smooth transition between them.

Finally, for different phenomenological approaches we wrote a linearized version of the equation (\ref{linear}). It takes into account interaction with only two gluons from the background fields and neglects all non-linear interactions. This equation can be applied in the region of intermediate $x_B$ and we believe will be very useful for analysis of EIC data.

The material presented here is based on paper \cite{mpaper} with I. Balitsky. I thank him for collaboration and guidance.

The author is grateful to G.A. Chirilli, J.C. Collins, Yu. Kovchegov,  M. D. Sievert, A. Prokudin,  A.V. Radyushkin, T. Rogers, and F. Yuan for valuable discussions. This work was supported by contract DE-AC05-06OR23177 under which the Jefferson Science Associates, LLC operate the Thomas Jefferson National Accelerator Facility.

\end{document}